# Leveraging the Flow of Collective Attention for Computational Communication Research

Cheng-Jun Wang*; Zhi-Cong Chen; Qiang Qin; Naipeng Chao*

Computational Communication Collaboratory, School of Journalism and Communication, Nanjing University, Nanjing 210093, P.R. China.

Correspondence and requests for materials should be addressed to CJW (wangchj04@126.com) or NC (npchao@nju.edu.cn)

Abstract: Human attention becomes an increasingly important resource for our understanding or collective human behaviors in the age of information explosion. To better understand the flow of collective attention, we construct the attention flow network using anonymous smartphone data of 100,000 users in a major city of China. In the constructed network, nodes are websites visited by users, and links denote the switch of users between two websites. We quantify the flow of collective attention by computing the flow network statistics, such as flow impact, flow dissipation, and flow distance. The findings reveal a strong concentration and fragmentation of collective attention for smartphone users, while the duplication of attention cross websites proves to be unfounded in mobile using. We further confirmed the law of dissipation and the allowmetric scaling of flow impact. Surprisingly, there is a centralized flow structure, suggesting that the website with large traffic can easily control the circulated collective attention. Additionally, we find that flow network analysis can effectively explain the page views and sale volume of products. Finally, we discuss the benefits and limitations of using the flow network analysis for computational communication research.

## Introduction

In the age of information explosion, human attention has become an increasing scarce resource (Goldhaber, 1997; Herbert A. Simon, 1973). Compared with the massive size of information, human attention is relatively limited. Thus, information resources tend to compete with each other to attract more attention. As a result, we observe the transportation of collective attention flowing from one to another. By tracing the switching of attention for individuals, researchers can track the flow of collective attention, and better understand the patterns as well as their potential impacts on human behaviors. To capture the flow of human attention can help us understand the patterns of attention dynamics, and it becomes an even bigger research question if we could find the connection between attention flow and the other kinds of human behaviors.

We probed the problem about how the collective attention flows in the perspective of computational social science. As an emerging paradigm of research, computational social science provides a powerful lens for researchers to understand human communication behaviors, with the increasing capability of collecting and analyzing very large scaled datasets. According to Watts (2007, p. 489), "data about Internet-based communication and interactivity and revolutionize our understanding of collective human behavior". It also provides rich theoretical perspectives, among which network science is very useful for investigating human interactions, because "we live life in the network" (Lazer, Pentland, Adamic, & Aral, 2009, p. 721), and network science provides a mathematical structure of many real social systems, including the World Wide Web, human society, and communication networks. Using the big data of individual browsing history, we can trace the successive attention flow between two information resources to build up the attention flow network and conduct the flow network analysis.

Using a novel dataset of 100,000 smartphone users in a major city of China, we investigated the patterns of the flow of collective attention. China has just entered the age of mobile Internet with the widely use of 4G wireless network, which gives rise to the rapid development of mobile phone applications, and has profound influence on the rapid social change in China and supply rich information about the flow of human attention. The findings reveal a strong concentration of collective attention allocated to popular websites for smartphone users, and strong scaling relationships for the flow of collection attention in the flow networks. Especially, there is a centralized flow structure, the website of large traffic can control the circulated collective attention. To investigate how flow network analysis can contribute to our understanding about the other kinds of human behaviors, we analyzed another dataset of online shopping. By constructing the attention flow network, we find that flow network analysis can effectively explain the sale volume of products.

In the perspective of metabolism, we consider information systems (e.g., websites) as "virtual living organisms that grow at the expense of users' attention" (L. Wu, Zhang, & Zhao, 2014, pp. 1-2), and they serve the users with massive information. First, we first reviewed prior studies on the flow of collective attention. Second, we introduced the study on flow network, and established our analytical framework focusing on three aspects, namely flow dissipation, flow impact, and flow distance. Third, our findings confirmed the well-established scaling patterns of collective attention, discovered the centralized allocation of collective attention for mobile phone use, and showed the potential of flow network analysis in collective human behavior. Finally, we also discussed the benefits and limitations of using the flow network analysis for computational communication research.

## The Flow of Collective Attention

Collective attention has been well studied in communication research in general and the studies of media choice in specific. Although human attention can be investigated on the individual level, only when the individual attention "scale up to form audiences that they are relevant to understanding the attention economy" (James G. Webster, 2014, p. 7). When users make up their decisions on media choice, the collective attention flows from one piece of information to another. Audience flow as one form of attention flow has been identified as a robust pattern in TV watching behaviors. People who watch one TV program would stay tuned to the next, leading to a disproportionately large amount of "duplicated audiences" as a result of "inheritance effects" (Headen, Klompmaker, & Rust, 1979; J. G. Webster, 1985).

There has been a large body of research on media choice, emphasizing on different aspects including rational choice (Steiner, 1952), genre preferences (Hsu, Özgecan, & Hannan, 2009; James, 2006), tastes (Bourdieu, 1984; Gans, 1999), use and gratifications (Alan, 1983; Katz, Blumer, & Gurevitch, 1974), selective exposure (Sears & Freedman, 1967). However, "people aren't always very good at matching their motives with media choices. Songs or programs in a preferred genre are missed, gratifications sought aren't obtained, and moods go unmanaged", which leads to the user's dilemma (James G. Webster, 2014, p. 35). Audiences are more featured by their bounded rationality (H. A. Simon, 1956), and therefore they heavily rely on repertoires (Ferguson & Perse, 1993; Heeter, 1985; Kimberly, David, & Leo, 2001) and heuristics (Metzger, Flanagin, & Medders, 2010) to manage their choices. Additionally, social networks and new media technologies also play important roles in the process of media choices. These driving forces of media choice shape the flow of collective attention.

Webster (2011) proposed a structurational theory of collective attention to explain how collective attention take shape in digital media environments. This structurational model includes three major constructs: users (agents), media (structure), and public measures (duality or structuration) (James G. Webster, 2014, p. 140). Two kinds of public measures are involved: the market information regimes (e.g., advertising, audience rating) push media to users, and the user information regimes (e.g., search and recommendation system) pull media to users. Through the duality or structuration process, users and media "mutually constitute the media environment" (James G. Webster, 2014, p. 136). However, as an exercise of data collecting, filtering, and reporting, public measures also bring with them potential biases (James G. Webster, 2011). Finally, the audiences take shape during the interplay between users and media.

"It is how audiences flow from one media encounter to the next that is of central interest" (James G. Webster, 2014, p. 66). The flow of collective attention supplies a new window for our understanding of audience fragmentation and social polarization. The competition for collective attention further "fragmented audiences across an ever-increasing number of outlets" (James G. Webster, 2014, p. 11). The distribution of collective attention attracted to each media resources can be well described as a long tail (media-centric approach). On the one hand, most users allocate their attention to a handful number of media. Therefore, there is strong concentration of collective attention. On the other hand, the development of digital media tends to supply more information, and makes the long tail a salient social issue. "Long Tail forces and technologies that are leading to an explosion of variety and abundant choice in the content we consume are also tending to lead us into tribal eddies. When mass culture breaks apart it doesn't reform into a different mass. Instead, it turns into millions of microcultures" (Anderson, 2006, p. 183).

Against the backdrop of media-centric approach, Webster and Ksiazek (2012) suggested to adopt the network approach to capture the fragmentation of collective attention. It aims to measure the extent to which audiences for multiple media outlets overlap with each other. And therefore it is also known as audience-centric fragmentation. In the constructed media network, nodes represent media outlets, and the link between any two nodes represents the common audiences between them. To capture the actual duplication of collective attention between two outlets, the links between two nodes whose observed duplication is less than the expected value would be removed. Using this method, Webster et al. (2012) reported that there is a strong overlapping of collective attention between two media outlets (e.g., tv channels or websites), which suggests that the fragmentation of collective attention may not produce highly polarized audiences. Based on the introduction above, we propose the following hypotheses:

*H1*: The flow of collective attention in the mobile information system is featured by strong concentration and fragmentation.

*H2*: There is a strong duplication of collective attention across media platforms in the mobile information system.

## Attention Flow Network

Network science provides a new perspective to analyze the flow of collective attention. By tracing the flow of individual attention, we can build up the flow network for collective attention. In the constructed network, nodes are the information resources visited by users, and edges represent the switching of collective attention from one node to another. Different from the other networks, the edges are both directed and weighted, and more importantly, the inflow must equal the outflow for each node in attention flow network. Therefore, we have to add two external nodes ("source" and "sink") to balance the flow of the nodes. The balanced flow network is also regarded as "open flow network", since it considers the flow from and to the environment (Guo et al., 2015). The characteristics of the attention flow networks discussed above calls for a new approach of analysis. The research on flow network analysis supplies a systematic way to understand the flow of collective attention.

We frame our analysis in the perspective metabolism, and the information system (e.g., websites) are viewed as "virtual living organisms that grow at the expense of users' attention" (L. Wu et al., 2014, pp. 1-2). The metabolism rate can be measured by the amount of energy needed per second to keep and organism alive. Generally speaking, a human being needs 2000 calories each day, corresponding to a rate about 90 watts. However, as West (2017) has introduced, nowadays human beings require homes, lighting, heating, transportation systems, and computers among others, in addition to the energy got from food. Thus, to capture the real metabolic rate, social metabolic rate can be defined based on the various energy resources human beings consume everyday. Similarly, we can also define the virtual metabolic rate for online social systems. To maintain its virtual life, online social systems also need to get the nutrients and energy from the outside world. The number of page views (*PV*) for an online social system can be viewed as the energy or nutrients, and the number of unique visitors (*UV*) of an online social system (e.g., a website) can be viewed as the "body mass". By absorbing users' attention and dissipate part of the attention out, website maintains its "life" and produces a massive amount of information to attract more attention, which gives rise to the allowmetric growth. Allometric growth describes the power law relationship between size and energy over time. For example, there exists a scaling relationship between *UV* and *PV*,

$$PV \sim UV^{\eta}. \tag{1}$$

By reviewing current research on flow network analysis, we outline our analytical framework, with particular focus on three aspects of the attention flow network, namely flow impact, flow distance, and flow dissipation.

**Flow Dissipation.** For any node $i$ in the attention flow network, the amount of attention flowing from $i$ to sink is defined as its dissipation, $D_i$. We can also capture the other similar measurements of the flow network. For example, the flow passing through node $i$ is denoted as $A_i$. The flow coming directly from source is denoted as $S_i$, and the flow out of node $i$ which does not flow directly to the sink is denoted as $F_i$.

For a balanced attention flow network at a given time point $t$,

$$UV_t = \sum_i^N D_i, \qquad (2)$$

$$PV_t = \sum_i^N A_i. \qquad (3)$$

Similar to the allowmetric scaling relationship between $PV$ and $UV$, there also exists a power law relationship between $D_i$ and $A_i$,

$$D_i \sim A_i^\alpha, \qquad (4)$$

where the scaling exponent $\alpha$ denotes the efficiency of dissipation. Eq (12) is also regarded as "the dissipation law" (Cheng-Jun Wang & Wu, 2016; Jiang Zhang & Wu, 2013) On the contrary, $1/\alpha$ can be used as an indicator of the attractiveness of the information resource (Cheng-Jun Wang & Wu, 2016).

If the dissipation exponent $\alpha$ is larger than 1, the flow dissipation from node $i$ increases faster than the flow through node $i$. The nodes with a larger flow tend to dissipate more flow to sink, which corresponds to the star-like flow network topology (Lingfei Wu & Zhang, 2013; Jiang Zhang & Wu, 2013). On the contrary, if the dissipation exponent $\alpha$ is smaller than 1, the nodes with a larger amount of traffic tend to dissipate less flow to the sink, which corresponds to the chain-like flow network topology (Lingfei Wu & Zhang, 2013; Jiang Zhang & Wu, 2013).

**Flow Impact.** The flow impact $C_i$ of a node in attention flow network measures its capability in controlling the flow circulation of the network. The idea of flow impact comes from the study of *Kleiber's law*. Max Kleiber (1947) found animals' metabolic rates scale to the 3/4 power of their body weights. Let $A$ denote the metabolism rate or the total flow from source to the network, and let $C$ represent the total mass or the summation of all individual flow rates in the network,

$$A \approx C^\gamma, \qquad (5)$$

and $\gamma$ is the allometric exponent. Interestingly, later research showed the allometric scaling relationship between $A$ and $C$ holds for the river networks, and $\gamma = 2/3$ in this case (Banavar, Maritan, & Rinaldo, 1999). Dreyer (2001) further created a *1d* space and ran water over permeable tissue. The relationship between the length of tissue wet by water $A$ and the total mass of water $C$ also follows the allowmetric scaling law, and the exponent $\gamma = 1/2$. Researchers tried to explain the allometric scaling law following the assumption that the most efficient transportation network was embedded in a *d* dimensional space, for example, the water through permeable tissue was embedded in a 1*d* space ($\gamma = 1/2$), the river network was embedded in a 2*d* space ($\gamma = 2/3$), and the vascular network embedded in a 3*d* space ($\gamma = 3/4$). Therefore, it seems that there exists an explicit relationship between $\gamma$ and *d*, $\gamma = d/(d+1)$ (Banavar et al., 1999; Dreyer, 2001; Jiang Zhang & Wu, 2013).

West et al. (1997, 1999) proposed that the *Kleiber's law* origins from the fractal structure of transportation system within living system, and they developed a model of spacefilling hierarchical networks to explain the observed quarter-law allometric scaling. Banavar et al. (1999) proposed a method to deal with the discrete transportation network, especially the river networks. They defined the size of the drainage basin of a river as *A* and the total amount of water contained in the water as *C*. For a sub-basin *X* of the river,

$$A_X = \sum_{Z \in nn(X)} A_Z + 1, \tag{6}$$

where *nn(X)* are the nearest neighbours that drains into *X*, and

$$C_X = \sum_{Z \in \gamma} A_Z, \tag{7}$$

in which $\gamma$ is "the collection of all sets connected to *X* through drainage directions" (Banavar et al., 1999, p. 132). Dreyer (2001) proposed a continuous approach similar to the method of Banavar et al. (1999). Inspired by the model of Banavar et al. (1999), Garlaschelli et al. (2003) proposed the method to compute *A* and *C* by converting the flow network to a minimal spanning tree. For the minimal spanning tree, $A_i$ of node *i* is the throughflow of the node, and $C_i$ is the total throughflow of the nodes rooted from node *i*. The method of Garlaschelli et al. (2003) has obvious shortcomings for flow network analysis: first, when some edges are removed, the information about the flow is lost; second, it can only be used for binary network rather than weighted networks.

Zhang and Guo (2010) proposed a systematic method to solve the problem: Given a flow network *G* with *N* nodes, two artificial nodes "source" and "sink" need to be added into the network to balance the inflow and the outflow of each node. Node *0* is the "source", and node *N+1* is the "sink". The balanced flow network can be expressed as a flow matrix *F*. The element $f_{ij}$ in *F* denotes the weight of flow from node *i* to *j*. The flow matrix *F* can be normalized by row to get the transition matrix *M*, and the element $m_{ij}$ of *M* satisfies

$$m_{ij} = f_{ij} / \sum_{k=1}^{N+1} f_{ik}. \tag{8}$$

The fundamental matrix *U* can be derived from *M*,

$$U = I + M + M^2 + \cdots = \sum_{i=0}^{\infty} M^i = (I - M)^{-1}, \tag{9}$$

in which *I* is the identity matrix. The element $u_{ij}$ in the fundamental matrix *U* represents the total flow from *i* to *j* along all possible pathways. Using the flow matrix *F*, we can obtain the total flow through any given node *i*,

$$A_i = \sum_{j=1}^{N+1} f_{ij}. \tag{10}$$

Accordingly, the flow impact $C_i$ can be defined as the following to capture the total direct and indirect flow stored in the sub-network rooted from node $i$,

$$C_i = \sum_{k=1}^{N} \sum_{j=1}^{N} \left(f_{0j} u_{ji}/u_{ii}\right) u_{ik}. \tag{11}$$

According to Zhang and Guo (2010), both $A_i$ and $C_i$ in the weighted food webs follow the power law distribution, and the relationship between $A_i$ and $C_i$ also follows a power law relationship,

$$C_i = A_i^{\eta}. \tag{12}$$

Garlaschelli et al. (2003) argues that for the binary network, $C_i$ represents the cost of the transportation, and the scaling exponent $\eta$ represent the efficiency of transportation. $\eta$ ranges from 1 to 2, where 1 denotes the most efficient network (star-like network), and 2 denotes the least efficient network (chain-like network). While Zhang and Guo (2010) interpret the meaning of $C_i$ and $\eta$ differently. $C_i$ is the total influence of node $i$ on the other nodes in the flow network, and $\eta$ represent the capability of the network to store flow within the system. The flow network with larger $\eta$ can store more flow "by means of cycling the flows in the network" (J. Zhang & Guo, 2010, p. 765). Additionally, the range of $\eta$ is not exactly [1, 2], and the range [1, 2] only applies to the minimal spanning tree (J. Zhang & Guo, 2010).

It is also necessary to note that $\eta$ also reflects the degree of centralization (Jiang Zhang & Wu, 2013). For an attention flow network whose flow passing through node $i$ is $A_i$, with the increasing of $\eta$, the distribution of $C_i$ can become much more uneven. The flow network with $\eta > 1$ is centralized, while the flow network with $\eta < 1$ is decentralized. Employing this method, it is found that most trade networks are centralized (P. T. Shi, Luo, Wang, & Zhang, 2013). In previous research, the flow impact exponent $\eta$ and dissipation exponent $\alpha$ were usually negatively correlated (L. Wu et al., 2014; Jiang Zhang & Wu, 2013) .

Wu and Zhang (2013) analyzed the flow structure of the clickstream on the Web. Clickstream is an ordered sequence of webpages visited by users. Using clicksteam as a proxy of the flow of collective attention, they confirmed the power law relationship between $C_i$ and $A_i$, but the scaling exponent $\eta$ is smaller than 1, implying a decentralized flow structure of the World Wide Web (WWW). The overall decentralized flow structure for the whole WWW is natural, since the people with the same language tend to visit the websites within the sub-community of the entire WWW. However, they found that the scaling exponents $\eta$ for the sub-communities of different languages are regularly smaller than 1. By conducting network permutations, they further found that the decentralized flow structure for the collective attention is related to "the topological structure of the clickstream network rather than the distribution of weights on clickstreams" (Lingfei Wu & Zhang, 2013, p. 5).

**Flow Distance.** The flow distance of a node measures how many steps a random walker takes to reach the node from source in the attention network (Guo et al., 2015; C. J. Wang, Wu, Zhang, & Janssen, 2016). We can assume that there are many random walkers or particles randomly walking on the flow network. The first-passage flow $\phi_{ij}$ from $i$ to $j$ is defined as the number of random walkers that reach $j$ at each time step for the first time after they have visited $i$. Accordingly, the average number of steps that these random walkers have taken is defined as the first-passage flow distance $l_{ij}$. Similarly, the total flow $\rho_{ij}$ from $i$ to $j$ is defined as the number of random walkers that arrived at node $j$ at each time step (no matter it is the first time or not) after they have visited $i$. And the average number of steps these random walkers take is defined as the total flow distance $t_{ij}$.

Let $p_{ij}^k$ denote the probability that random walkers transfer from $i$ to $j$ after $k$ steps. Then the total flow distance $t_{ij}$ from $i$ to $j$ along all possible pathways can be obtained,

$$t_{ij} = \sum_{k=1}^{\infty} k p_{ij}^k. \tag{13}$$

Given the flow from $i$ to $j$ after k steps is $\phi_{0i}(M^k)_{ij}$ and the total flow from $i$ to $j$ is $\rho_{ij}$, we have

$$p_{ij}^k = \phi_{0i}(M^k)_{ij}/\rho_{ij}. \tag{14}$$

Based on Eq. (13) and (14), Guo et al. (2015) get the expression for $t_{ij}$:

$$t_{ij} = \sum_{k=1}^{\infty} k \frac{\phi_{0i}(M^k)_{ij}}{\rho_{ij}} = \frac{\phi_{0i}\left(\sum_{k=1}^{\infty} k M^k\right)_{ij}}{\rho_{ij}} = \frac{\phi_{0i}(MU^2)_{ij}}{\rho_{ij}} = \frac{\phi_{0i}(MU^2)_{ij}}{\phi_{0i}u_{ij}} = \frac{(MU^2)_{ij}}{u_{ij}}, \tag{15}$$

where $M$ is the transition matrix, and $U$ is the aforementioned fundamental matrix, and $u_{ij}$ is an element of $U$.

Similarly, the expression for the first-passage flow distance can be obtained. Let $q_{ij}^k$ denote the probability that random walkers jump from $i$ to $j$ after $k$ steps for the first time. Then the first-passage flow distance $l_{ij}$ from $i$ to $j$ along all possible pathways can be expressed,

$$l_{ij} = \sum_{k=1}^{\infty} k q_{ij}^k. \tag{16}$$

Given the expression for the probability $q_{ij}^k = \phi_{0i}(M_{-j}^k)_{ij}/\phi_{ij}$, Guo et al. (2015) showed that $l_{ij}$ can further be expressed,

$$l_{ij} = t_{ij} - t_{jj} = \frac{(MU^2)_{ij}}{u_{ij}} - \frac{(MU^2)_{jj}}{u_{jj}}. \tag{17}$$

Based on first-passage flow distance $l_{ij}$ and $l_{ji}$, the symmetric flow distance $c_{ij}$ can be defined,

$$c_{ij} = 2l_{ij}l_{ji}/(l_{ij} + l_{ji}). \tag{18}$$

Flow network can be embedded into a Euclidean space in which the distance between any two nodes equals their flow distance. Flow distance reflects the nature about how random walkers move on the flow network, thus it connects the network topology with flow dynamics. By embedding flow network into a proper geometric space, our understanding about the underlying dynamics may be simplified. For example, using the effective distance instead of conventional geometric distance, Brockmann and Helbing (2013) captured the hidden geometry of the air-traffic-mediated epidemics, and found a very good linear relationship between the arrival times for cities and the effective distance.

Shi et al. (2015) further specified the network embeding method based on symmetric flow distance $c_{ij}$, and applied the method to the clicksteam network of Indiana Unversity. Their findings indicated that websites can be grouped intro three layers (P. T. Shi et al., 2013). The most popular websites, such as google.com, are in the central of the embedded geometric space featured by a large amount of attention flow and a small fraction of dissipation. A large number of websites with a small amount of attention flow and a large fraction of dissipation lies between the core and the periphery. And the small websites with few attention flow and a small fraction of dissipation stays in the periphery.

Flow distance is important for us to understand the lifecycle of news. For example, Wang et al. (2016) measures the flow distance for the news on *Digg*—a social news website, and they found that attention flow network of news stories preserved a stable tree-like structure over time which led to the scaling between the number of users and clicks.

Based on the introduction about flow networks, we deprive the following hypotheses:

*H3*: There are strong scaling relationships for the collective attention flow in mobile information system. To be specific:

*H3a*: There exists a law of dissipation for the collective attention flow in mobile information system.

*H3b*: There exists a law of flow impact for the collective attention flow in mobile information system.

*H4*: The flow network statistics, such as flow dissipation, flow impact, flow distance can help predict the over all page views as well as sale volumes.

## Method

A novel dataset of 100,000 smartphone users in a major city of China was used to investigate the patterns of the flow of collective attention. To test how flow network analysis can help explain the sale volume of products, we analyzed another dataset of online shopping. To better understand the flow of collective attenion, we constructed the attention flow network (Cheng-Jun Wang & Wu, 2016). In the attention flow network, nodes represent websites visited by users, and links are the switching of users between two websites. There are 25,753 nodes, and 376,118 links in this attention flow network.

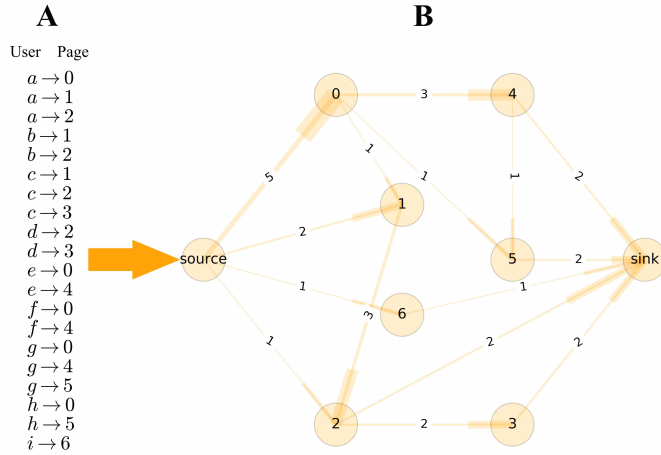

Figure 1. Constructing the Flow Network

To balance the flow into a node (in-flow) and the flow out of a node (out-flow), we add two external nodes—source and sink. See *Figure 1* for example, panel A shows the edgelist data format we use to build up the network. There are two columns, the first column denotes users, and the second column denotes pages. For a node *i* whose out-flow is larger than its in-flow, we add a directed weighted edge from source to node *i*; for a node *j* whose out-flow is smaller than its in-flow, we add directed weighted edge from node *j* to sink. The python package flownetwork has been developed for the data analysis of flow networks (Cheng-Jun Wang, 2017).

After the construction of attention flow network, we calculate the flow impact, flow dissipation, and flow distance. We first explore the patterns of the observed attention flow network, especially the scaling patterns for flow impact, flow dissipation. Then we tried to investigate how can the flow network statistics help understand the allocation of human attention and the sale volume of products using OLS regression models.

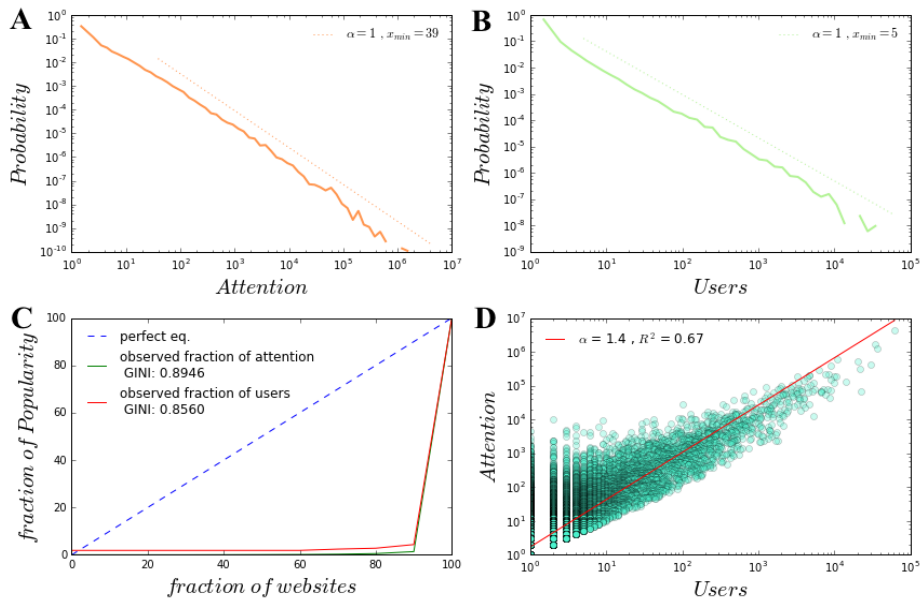

Figure 2. Concentration of Collective Attention

**Findings**

First, we find strong attention concentrations for the smartphone users. The popularity of individual website can be measured by either the unique number of users who have visited the website (users, UV) or the total of time spend by these users (attention, PV). In Figure 2, Panel A and Panel B demonstrates the long tail distribution of the popularity of websites measured by attention and users respectively; Panel C demonstrates the gini coefficients for the popularity of attention (Gini coefficient = 0.89) and the popularity of users (Gini coefficient = 0.86); Panel D shows the superlinear scaling relationship between two measures of popularity (alpha = 1.4). Thus, the websites who are more successful to attract users, can also get more attention. Using such media-centric approach, the long tail distribution in Figure 2 also reveals that there is a strong fragmentation of collective attention (James G. Webster, 2014; James G. Webster & Ksiazek, 2012). Therefore, *H1* is well supported.

Second, we investigated the duplication of collective attention following the approach proposed by Webster and Ksiazek (2012). However, as Figure 3A shows, the cross-platform degree score also follows a long tail distribution. Against the equality found by Webster and Ksiazek (2012), the duplication of collective attention is very unequal. Comparing the original Zipf's distribution with the revised Zipf's distribution, we found that removing the edges following the approach suggested by Webster and Ksiazek (2012) can not make substantial difference from the original network (See Figure 3A). As we have introduced, the flow impact $C_i$ captures the total direct and indirect flow stored in the sub-network rooted from node *i*. In other words, flow impact also measures the "duplication" between node *i* and the sub-network rooted from it. Figure 3B demonstrates that the distribution of flow impact also largely follows a long tail distribution. *H2* is not supported.

Figure 3. Duplication of Collective Attention

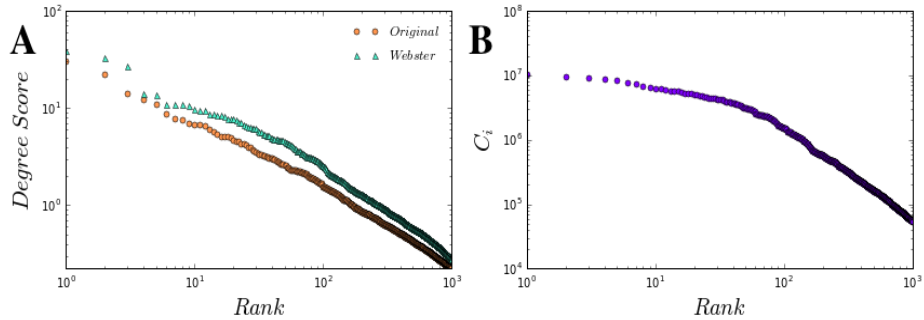

Third, we find strong scaling patterns for the attention flow network (see Figure 4). As we have introduced above, for node *i*, the flow through the node, the dissipation to the sink, the flow directly from source, and the flow coming out of node *i* and flowing to the other nodes except of sink are denoted as $A_i$, $D_i$, $S_i$, and $F_i$, respectively. The findings confirmed the scaling relationship between $D_i$ and $A_i$, $D_i \approx A_i^{\alpha}$, and the power exponent ($\alpha$ = 0.79) measures the capability of the information system to keep collective attention. Similarly, we can also quantify the capability of the attention flow network to attract attention form outside of the information system, $A_i \approx S_i^{\beta}$, $\beta = 1.44$. Thus, *H3a* is well supported.

Figure 4. Scaling of Attention Flow

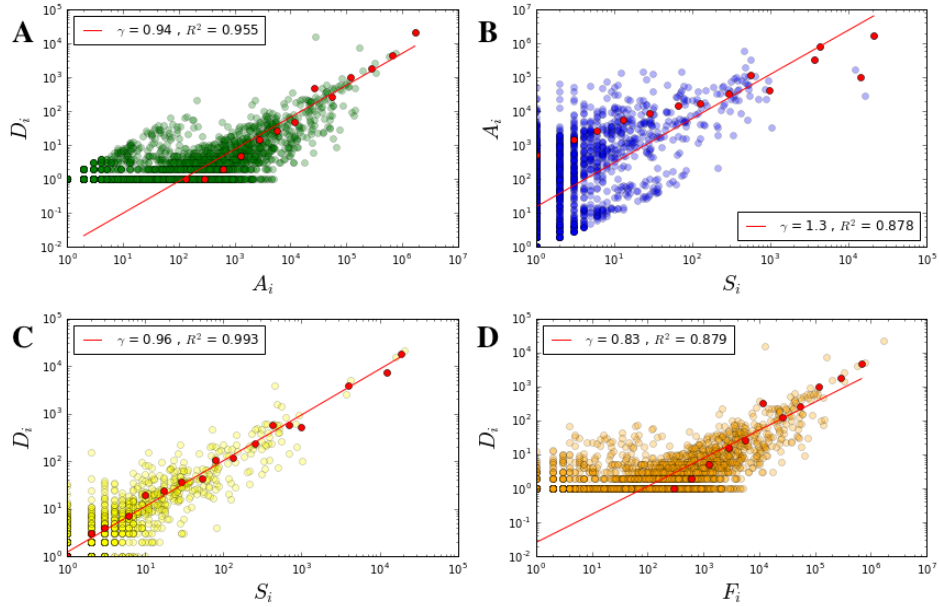

Fourth, the flow impact $C_i$ in the attention flow network can be quantified by the circulated traffic it controls (Lingfei Wu & Zhang, 2013). The scaling relationship between $C_i$ and $A_i$ is well captured by the dissipation law, $C_i \approx A_i^\eta$. $\eta = 1.12$ (Figure 5). $\eta$ measures the capability of the system to store collective attention, and $\eta > 1$ indicates that the information system of the mobile phone has a strong capability to maintain human attention. H3b is also well supported.

Figure 5. Scaling Relationship Between Impact and Traffic

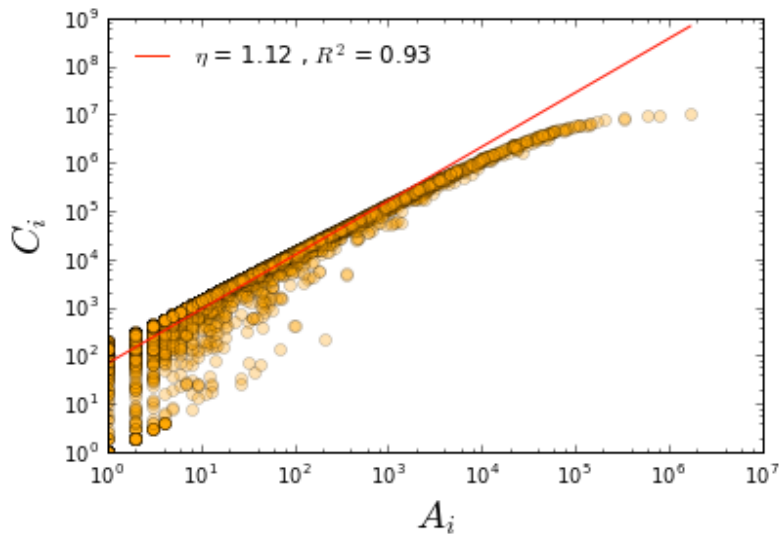

Finally, we further employ the OLS regression models to investigate how flow network statistics can help understand the allocation of collective attention and the sale volume of products. As model 1 in *table 1* demonstrates, the flow statistics can effectively explain the variance of *PV* ($R^2=.924$). The regression coefficients of $D_i$, $S_i$, $C_i$, $l_{ij}$ are .679, -.344, .856, and -.064, respectively, and the effects are significant. Using another dataset of taobao, we also build up the attention flow network, calculated the same flow network statistics ($D_i$, $S_i$, $C_i$, $l_{ij}$), and construct the OLS regression to explain the sale volume of products (see model 2 in Table 1). The R square is .226, and all the regression coefficients of $D_i$, $S_i$, $C_i$, $l_{ij}$ are significant. In all, the flow network statistics can effectively explain the page views and sale volume. Therefore, *H4* is supported in our research.

Table 1. OLS Regression on Page Views and Sale Volume

|  | Model 1 | Model 2 |
| --- | --- | --- |
|  | Page Views (log) | Sale Volume (log) |
| *Constant* | -1.183*** (.026) | .264*** (.025) |
| $D_i$ *(log)* | .679*** (.012) | 1.293*** (.028) |
| $S_i$ *(log)* | -.344*** (.013) | .476*** (.031) |
| $C_i$ *(log)* | .856*** (.002) | .585*** (.010) |
| $l_{ij}$ | -.063*** (.001) | -.045*** (.001) |
| $R^2$ | .924 | .226 |

**Conclusions and Discussions**

Digital traces of human behaviors supply precious information about how human beings allocate their increasingly limited attention to various information resources. In the perspective of metabolism, online information system can be viewed as a virtual living organism. It lives on the collective attention and yields tremendous amount of information to its users. The flow of collective attention shapes the information system. Similar to the study on ecosystem, previous research has already found the allometric growth of online information system by measuring the evolution of *PV* and *UV* over time (Cheng-Jun Wang & Wu, 2016). Our research is based on this general framework, and our data analysis of attention flow network confirms various scaling relationships for flow impact, inflow, flow dissipation, flow distance, *PV*, and *UV*.

The long tail distribution suggests that the collective attention circulated on the 4G mobile Internet is both concentrated and fragmented. Using the method proposed by Webster and Ksiazek (2012), we find very little duplication across media outlets. Flow impact captures the potential for the users who visited a website will keep moving forward and visiting the other websites. Since measurement of flow impact considers not only the direct flow but also the indirect flow, it can better reflect the duplication of collective attention. Although the heterogeneity of flow impact is relatively smaller for the "head" than the "long tail", the Zipf's distribution of flow impact is also largely featured by a long tail. Thus, we did not find a strong duplication of collective attention. The fear for attention polarization is not groundless especially for the mobile media.

The dissipation exponent $\alpha$ for the power law relationship between $D_i$ and $A_i$ is smaller than 1 ($\alpha = .79$, see Figure 2A). $1/\alpha$ can be used to measure the overall attractiveness of the informaiton system. Whe $\alpha$ is smaller than 1, the attractiveness of the popular nodes with a larger amount of traffic is relatively large, and they tend to dissipate less flow to the sink. Furtehr, it suggests that the attention flow of the smartphone phone user tends to be a chain-like network topology. This finding supports the idea that there exists obvious hierachy structure in the obseved attention flow network (P. Shi et al., 2015).

For the clickstream flow network of WWW, the scaling exponent $\alpha$ for the power law relationship between $C_i$ and $A_i$ is smaller than 1, implying a weak ability for WWW to maintain the collective attention (L. Wu et al., 2014); while for the attention flow network of 4G mobile Internet, we find that the scaling exponent $\eta$ is larger than 1. It suggests that the information system of the mobile phone has strong capability to attract people to stay in the system. Also, when $\eta > 1$, it implies that the website with large traffic can more strongly dominate the circulation of collective attention. Therefore, there is a centralized attention flow structure, which is compatible with the idea that there is a chain-like hierarchy topology. However, we also notice that when $A_i$ is larger than 10,000, the scaling relationship between $C_i$ and $A_i$ get weak (see Figure 4), which may imply a decentralized structure among the 'super' nodes with large user traffic.

The regression analysis shows that dissipation and flow impact have a positive effect on both page views of websites and sale volume of products, while flow distance has a negative effect. It is intuitive to understand that the information resource of strong flow impact and short flow distances can achieve more success. However, the positive effect of dissipation on attention allocation and online purchasing behaviors may be ignored. We assert that larger dissipation is the signature of the quality of that piece of information resource. If it could satisfy the need of users, the users have no need to surf online anymore. For example, in the case of online shopping, if most users like a product and buy it immediately after they find it, the dissipation of the information about this product must be large. Or the information takes a lot of energy for the users to read through, and the users may get tired and leave the information system. In the case of online shopping, users may find a product of interest, but it takes a lot of time to read the detailed information about the product and comments of the other users to make up a decision. Therefore, the users get tired and leave the information system.

Using flow network analysis as a computational framework for communication research has many benefits. First, there have already been many seminal studies devoted to gauge various flow systems, including energy flow system (e.g., food web) (Garlaschelli et al., 2003; G. B. West et al., 1997, 1999; J. Zhang & Guo, 2010), and river networks (A.-L. Barabási & Albert, 1999), Therefore, there have already been many theoretical perspectives for our research on attention flow networks (Cheng-Jun Wang & Wu, 2016; C. J. Wang et al., 2016; Lingfei Wu & Zhang, 2013; L. Wu et al., 2014). Second, a lot of human communication behaviors can be represented as flow networks. We can apply flow network analysis to study how people watch videos, read news, switch between different mobile apps, etc. Third, the computational methods for analyzing the flow networks have been well developed. For example, a python package *flownetwork* has been developed for flow network analysis (Cheng-Jun Wang, 2017). Third, the flow network analysis can be used to analyze the complete sequence of behaviors for a group of people. It is usually difficult to trace the information diffusion within an information system (e.g., Twitter), but it is easier to investigate how the users of a sub-community (e.g., journalists) allocate their attention over a series of information. Of course, there are also some limitations for the state of the art of flow network analysis. For example, it has strict requirements about the data. There should be complete records of behaviors for a group of people with explicit sequential information. And the computational complexity is still large, further optimization of the algorithms used in flow network analysis is necessary (J. Zhang & Guo, 2010).

"With the capacity of collecting and analyzing massive amount of data" (Lazer et al., 2009, p. 721), computational communication research is occurring. We define computational communication research as a branch of computational social science, with a particular focus on answering the important questions of human communication with the aid of big data and computational methods. Take this present research as an example, we focus on the salient question about the flow of collective attention, inspired by the development of computational social science (Lazer et al., 2009; D. J. Watts, 2007; Duncan J. Watts, 2013, 2017). Our daily life is increasingly networked and digitalized, the size and the form of the media evolves very fast. Drawing on the theoretical perspective of biology, especially the study of the ecology (Kleiber, 1947; G. B. West et al., 1997, 1999; L. Wu et al., 2014; J. Zhang & Guo, 2010), we believe that viewing the interaction between collective attention and information resources as a process of metabolism supplies a proper theoretical framework for our study. We also tried to leverage the power of big data of human behaviors to investigate the research question we aimed in solving. Two novel datasets about the human browsing behaviors are employed to construct the attention flow network. With the aid of novel data, salient question, and insightful theory, we explored the patterns and laws beneath the flow of collective attention. The strong regularities of human communication behaviors calls for the research to study the underlying mechanism or even the general principle governing human behaviors (A. L. Barabási, 2005; Huberman, Plt, Pitkow, & Lukose, 1998).